\documentclass[aip,jap,reprint]{revtex4-1}

\usepackage{soul}
\usepackage{bm}
\usepackage{xcolor}
\usepackage{graphicx}
\usepackage{amsmath}
\usepackage{booktabs}
\usepackage{appendix}
\usepackage{multirow}
\usepackage[english]{babel}
\usepackage{multirow}
\usepackage{colortbl}

\draft 

\begin{document}

\title{Exploring transport mechanisms in atomic precision advanced manufacturing enabled pn junctions}

\author{Juan P. Mendez}
\email[]{jpmende@sandia.gov}
\affiliation{Sandia National Laboratories, 1515 Eubank SE, Albuquerque, NM 87123, USA}
\author{Xujiao Gao}
\email[]{xngao@sandia.gov}
\affiliation{Sandia National Laboratories, 1515 Eubank SE, Albuquerque, NM 87123, USA}
\author{Jeffrey A. Ivie}
\affiliation{Sandia National Laboratories, 1515 Eubank SE, Albuquerque, NM 87123, USA}
\author{James H. G. Owen}
\affiliation{Zyvex Labs LLC, 1301 N. Plano RD, Richardson, TX 75081, USA}
\author{Wiley P. Kirk}
\affiliation{3D Epitaxial Technologies, LLC, 999 E. Arapaho Rd, Richardson, TX 75081, USA}
\author{John N. Randall}
\affiliation{Zyvex Labs LLC, 1301 N. Plano RD, Richardson, TX 75081, USA}
\author{Shashank Misra}
\email[]{smisra@sandia.gov}
\affiliation{Sandia National Laboratories, 1515 Eubank SE, Albuquerque, NM 87123, USA}

\date{\today}

\begin{abstract}
We investigate the different transport mechanisms that can occur in pn junction devices made using atomic precision advanced manufacturing (APAM) at temperatures ranging from cryogenic to room temperature. We first elucidate the potential cause of the anomalous behavior observed in the forward-bias response of these devices in recent cryogenic temperature measurements, which deviates from the theoretical response of a silicon Esaki diode. These anomalous behaviors include current suppression at low voltages in the forward-bias response and a much lower valley voltage at cryogenic temperatures than theoretically expected for a silicon diode. To investigate the potential causes of these anomalies, we studied the effects of a few possible transport mechanisms, including band-to-band tunneling, band gap narrowing, potential impact of non-Ohmic contacts, band quantization, impact of leakage, and inelastic trap-assisted tunneling, through semi-classical simulations. We find that a combination of two sets of band-to-band tunneling (BTBT) parameters can qualitatively approximate the shape of the tunneling current at low bias. This can arise from band quantization and realignment due to the strong potential confinement in $\delta$-layers. We also find that the lower-than-theoretically-expected valley voltage can be attributed to modifications in the electronic band structure within the $\delta$-layer regions, leading to a significant band-gap narrowing induced by the high density of dopants. Finally, we extend our analyses to room temperature operation and predict that trap-assisted tunneling (TAT) facilitated by phonon interactions may become significant, leading to a complex superposition of BTBT and TAT transport mechanisms in the electrical measurements.
\end{abstract}


\maketitle 

\section{Introduction}\label{sec:introduction}

Atomic precision advanced manufacturing (APAM) technology enables the creation of 2D doped regions, also known as $\delta$-layers, in semiconductors with single-atom precision \cite{Wilson:2006, Warschkow:2016, Fuechsle:2012, Ivie:2021b, Wyrick:2022} and high conductivity \cite{Goh:2006, Weber:2012, McKibbin:2013, Keizer:2015, Skeren:2020, Dwyer:2021}. APAM has various applications, including the exploration of novel electronic devices for classical computing and sensing systems \cite{Mahapatra:2011, House:2014, Skeren:2020, Gao:2021.2,Donnelly:2021}, or the exploration of dopant-based qubits in silicon, with recent advancements in understanding the advantages of leveraging the number of dopants as a degree of freedom \cite{Krauth:2022,Fricke:2021}, or the exploration of many body \cite{Wang:2022} and topological \cite {Kiczynski:2022} effects in dopant chains. To date, nearly all of these devices have been made with donors \cite{Mahapatra:2011,House:2014,Donnelly:2021,Fricke:2021,Wang:2022,Kiczynski:2022}. Recently, Ref.~\onlinecite{Skeren:2020} reported the fabrication of the first APAM bipolar device, composed of a boron $\delta$-layer and a phosphorus $\delta$-layer embedded in silicon. Interestingly, the characteristic response of the device deviates from the theoretically expected behavior of a Si Esaki diode.

\begin{figure}[ht]
  \centering
  \includegraphics[width=0.95\linewidth]{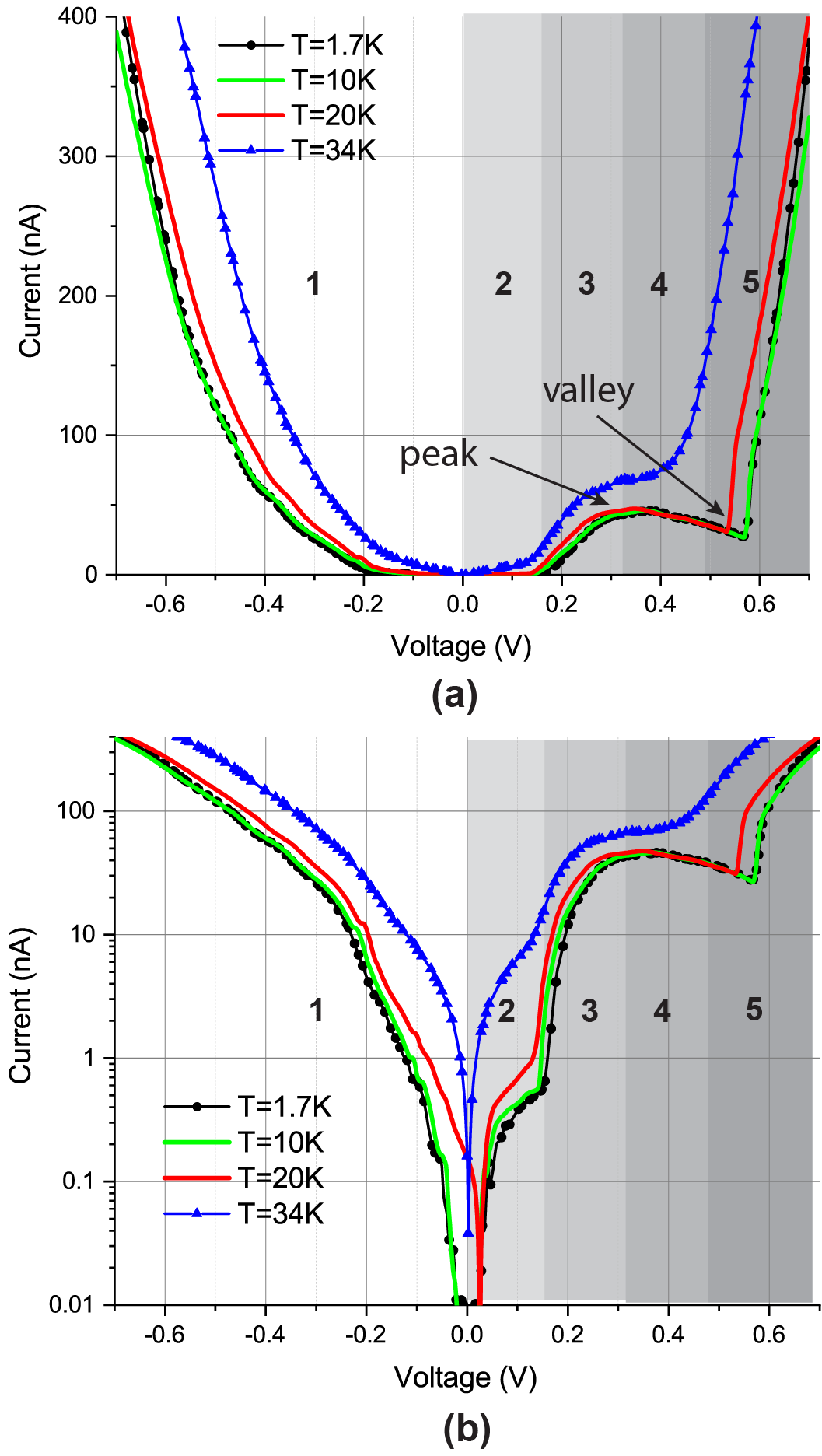}
  \caption{Measured current-voltage curves published in Ref.~\onlinecite{Skeren:2020}: \textbf{(a)} linear plot and \textbf{(b)} semilogarithmic plot. Note that the measured current, as shown in \textbf{(a)} and \textbf{(b)}, represents the absolute current, and the width of the pn device is 100~nm. The measured data was obtained from the repository cited in Ref.~\onlinecite{skeren_2020_data}.}
  \label{fig:Measured I vs V curve}
\end{figure}

Fig.~\ref{fig:Measured I vs V curve} shows the measured current-voltage (I-V) curves in both forward and reverse operation regimes at different temperatures. The authors of the paper speculated on the main transport mechanisms for these regions: (i) region 1 is due to increased band-to-band tunneling (BTBT) (a.k.a. Zener tunneling) under reverse voltage bias; (ii) region 2 may be due to band misalignment between the two $\delta$-layers, non-ideal contact effect, or localization effect in the boron $\delta$-layer; (iii) region 3 is due to increased BTBT under forward voltage bias; (iv) region 4 is due to decreased BTBT under forward bias; (v) region 5 is dominated by the above-threshold diffusion current as in a normal pn diode. We also note that, in region 5, the diode operates as a resistor, resulting in a linear current-voltage response, which appears as a linear curve in (a) and as a flatter curve on the logarithmic current scale in (b). In the 1.7-20~K range, the device response exhibits minimal temperature dependence. Conversely, above 20~K, the I-V measurements show a temperature dependence. As the authors in Ref.~\onlinecite{Skeren:2020} argued, at temperatures above 20-34~K, the bulk conductivity of the substrate surrounding  the pn device becomes significant, leading to considerable current leakage through this region, and thus an overall increase of the total current and the disappearance of the negative differential conductance (NDC).

From the results of Fig.~\ref{fig:Measured I vs V curve}, we can distinguish two qualitative differences between the measured data and the expected theoretical behavior from a Si Esaki diode - the suppression of BTBT in region 2\cite{Chynoweth:1961,Oehme:2009,Schmid:2012} and the relatively low voltage for the onset of region 5 (below 0.6~V) compared to the theoretical value (approximately around 1.0~V) in silicon diodes at cryogenic temperatures. 

In this work, we use semi-classical Technology Computer Aided Design (TCAD) to model different transport mechanisms expected for this structure, specifically to identify whether the unusual features in the I-V curves can be traced to the novel electronic properties of $\delta$-layers. We have found that the measured low-voltage onset in region 5 can be explained by a significant band gap narrowing induced by the high doping densities in the $\delta$-layer regions. We have also found that a combination of two sets of BTBT parameters can qualitatively approximate the shape of the tunneling current at low bias. This sheds light on a plausible cause of the atypical behavior, which is attributed to the quantization of the valence and conduction band structures in the p-type and n-type $\delta$-layers, respectively, due to the strong confinement of the dopants in one direction. This intriguing result may also be relevant for other tunnel diode systems, such as those involving 2D materials or van der Waals materials, where a high density of electrons and holes are strongly confined in 2D planes. We have also shown through simulations the possible existence of leakage current at temperatures above 30~K through the silicon enclosing the pn device, as observed experimentally, however we will exclude these leakage pathways in our simulations. Finally, we further extend the TCAD models to include inelastic trap-assisted tunneling (TAT) at room temperature. We have found that TAT may become significant at room temperature, leading to a complex superposition of BTBT and TAT transport mechanisms in the electrical measurements.

In the following, we describe our simulation approach in Sec.~\ref{sec: simulation approach}, explore major transport mechanisms across the $\delta$-layer-based pn junction and compare our results with available experimental data in Sec.~\ref{sec: transport mechanism}, and summarize our findings in Sec.~\ref{sec:conclusions}. 

\section{Simulation Approach}\label{sec: simulation approach}

In this work, we used the open-source TCAD code, Charon \cite{charon-website}, which is a multi-dimensional, MPI(Message Passing Interface)-parallel, semi-classical device simulation code developed at Sandia National Laboratories. It is commonly argued that quantum mechanical modeling \cite{Mamaluy:2021} is needed to accurately model APAM devices due to the sharp potential confinement in the $\delta$-layer. However, recent work \cite{Gao:2020,Gao:2021.2} shows that semi-classical Charon simulations can capture key current-voltage characteristics in APAM devices. For example, Ref.~\onlinecite{Gao:2020} shows that Charon simulations reproduce the measured sheet resistances very well when proper mobility values are used in phosphorous $\delta$-layers. Similarly, Ref.~\onlinecite{Gao:2021.2} shows that Charon simulations can predict the expected transistor-like current-voltage response in a proposed APAM vertical tunneling field effect transistor (v-TFET). In addition, it is well-known that semi-classical TCAD simulations run much faster than full quantum simulations and can be used to simulate complex device geometries. 

Sandia's TCAD Charon code solves the Poisson equation, the continuity equations for electrons/holes/ions, and the lattice temperature equation. These partial differential equations can be solved individually or coupled together, using either finite element \cite{Brooks:1982,Bochev:2013} or finite volume discretization schemes \cite{Bochev:2013b}.
The device geometries and meshes are generated using the tool called Cubit \cite{Cubit}. The simulations are visualized using the open-source data analysis and visualization tool called Paraview \cite{Paraview}.

Charon contains many advanced material models such as BTBT and trap-assisted tunneling (TAT) (refer to Appendix~\ref{appendix: TAT model} for the TAT model). In Charon, BTBT is modeled as a field-dependent generation rate which is given by \cite{Hurkx:1992, Wong:2020}
\begin{equation}
G_{btbt} = \alpha A \left( \frac{|F-F_{eq}|}{F_{eq}} \right)^\beta \left( \frac{F}{F_0} \right)^\gamma \textrm{exp} \left(-\frac{B}{F} \right). 
\label{eq:modified-b2bt}
\end{equation}
where $G_{btbt}$ is in unit of cm$^{-3}$s$^{-1}$, $F$ is the electric field magnitude in V/cm, and $\gamma$, $\beta$, $A$ and $B$ are fitting parameters. It was found in Ref.~\onlinecite{Hurkx:1992} that $\gamma = 2$ for direct transitions and $\gamma = 2.5$ for indirect transitions, namely, phonon-assisted band-to-band tunneling. $F_0$ is equal to 1~V/cm. The value of $\alpha$ can be -1, or 1. When $\alpha = -1$, we have $G_{btbt} < 0$, which indicates that BTBT is a recombination process. When $\alpha = 1$, we have $G_{btbt} > 0$, which indicates that BTBT is a generation process. $F_{eq}$ is the simulated, position-dependent electric field magnitude in a device under zero-current equilibrium condition. The field factor, $\left( |F-F_{eq}| / F_{eq} \right)^\beta $, theoretically guarantees zero current in equilibrium condition. $A$ and $B$ parameters are related to material properties such as effective masses and band-gap energy\cite{Kao:2012,Kane:1960,Vandenberghe:2010}, but they are often used as fitting parameters. These parameters were expressed for direct tunneling in Refs.~\onlinecite{Kao:2012,Kane:1960,Vandenberghe:2010} as
\begin{equation}
A= \bigg( \frac{g m_r^{1/2} (qF_0)^2}{\pi h^2 E_g^{1/2}} \bigg)
\label{eq:expression for A}
\end{equation}
and
\begin{equation}
B=\bigg( \frac{ \pi^2 m_r^{1/2} E_g^{3/2} }{qh} \bigg),
\label{eq:expression for B}
\end{equation}
where $g$ is a degenerate factor, $m_r$ is the reduced tunneling mass, $q$ is the elementary charge, $h$ is the Planck's constant and $E_g$ is the band gap.

Silicon is known to be an indirect band gap semiconductor material. However, it has been shown experimentally \cite{Miwa:2013,Miwa:2014,Mazzola:2018,Holt:2020,Mazzola:2020,Skeren:2020} and theoretically \cite{Carter:2009,Carter:2011,Lee:2011,Drumm:2013,Campbell:2023} that silicon doped with extreme doses of P, beyond the solubility limit, becomes a direct band gap material. Thus, for our calculations, we opted for a $\gamma$ value of $2.0$, substantiated by the existence of direct band-to-band tunneling in the measured IV in Ref.~\onlinecite{Skeren:2020} at very low temperatures; indirect band-to-band tunneling would be minimal at very low temperatures due to the absence of sufficient phonon-assisted tunneling processes. 

Additionally, for the simulations, we utilize the Slotboom model for intrinsic density and band gap narrowing \cite{Slotboom:1977}, alongside the mid-gap Shockley-Read-Hall (SRH) and Auger recombination models. For mobilities, we note that the Arora model\cite{Arora:1982} overestimates electron and hole mobilities for very high doping densities, particularly at very low temperatures, compared to the measured mobilities in $\delta$-layers~\cite{Weir:1994,Goh:2006,Ma:2012,Skeren:2020} (see Table~\ref{tab:Arora mobilities, BG and Electron affinity}). Therefore, in this work, we used the measured mobility values, indicated in Table~\ref{tab:Arora mobilities, BG and Electron affinity}, for the $\delta$-layer regions and the mobilities predicted by the Arora model for the regions with low doping densities. For more details of these models, we refer to the Charon's manual \cite{TCADCharon}. 

\begin{table*}[ht]
\begin{ruledtabular}
\begin{tabular}{l l l l l l l l}
\toprule
Temperature (K)                                         &                                               & 4         & 30      & 50      & 75      & 100     & 300    \\ 
\hline
\multirow{4}{*}{$\mu_e$(cm$^2$V$^{-1}$s$^{-1}$)}  & Arora\cite{Arora:1982}                        & 1031.0    & 327.0   & 244.4   & 194.1   & 165.0   & 89.3  \\
                                                  & Ref.~\onlinecite{Goh:2006,Goh:2009}           & 34-56     &         &         &         &         &        \\
                                                  & Ref.~\onlinecite{Ma:2012}                     &           &         &         &         &         & 25-40  \\
                                                  & Used in this work                             & 40        &         &         &         &         & 30     \\
\arrayrulecolor{gray}\hline                                                  
\multirow{4}{*}{$\mu_h$(cm$^2$V$^{-1}$s$^{-1}$)}  & Arora\cite{Arora:1982}                        & 636.2     & 201.8   & 150.8   & 119.8   & 101.8   & 55.1  \\
                                                  & Ref.~\onlinecite{Skeren:2020}                 & 13.6-19.0 &         &         &         &         &        \\
                                                  & Ref.~\onlinecite{Weir:1994}                   & 24        & 24.3    & 24.3    & 24.0    & 23.6    & 18.6   \\
                                                  & Used in this work                             & 20        &         &         &         &         & 18.6   \\
\arrayrulecolor{gray}\hline                                                  
$Eg$ (eV)                                         & \multirow{2}{*}{Slotboom\cite{Slotboom:1977}} & 0.9815    & 0.9808  & 0.9798  & 0.9777  & 0.9751  & 0.9360  \\
$\chi$ (eV)                                       &                                               & 4.2195    & 4.2198  & 4.2203  & 4.2213  & 4.2227  & 4.2422 \\
\bottomrule
\end{tabular}
\end{ruledtabular}
\centering
\caption{Electron ($\mu_e$) and hole mobilities ($\mu_h$) from experimental measurements\cite{Weir:1994,Goh:2006,Goh:2009,Skeren:2020} and those predicted by the Arora model\cite{Arora:1982}, as well as the effective band gap ($E_g$) and electron affinity ($\chi$) predicted by the Slotboom model\cite{Slotboom:1977}, for a doping density of $3 \times 10^{20}$~cm$^{-3}$.}
\label{tab:Arora mobilities, BG and Electron affinity}
\end{table*}

\begin{figure}[ht]
  \centering
  \includegraphics[width=0.95\linewidth]{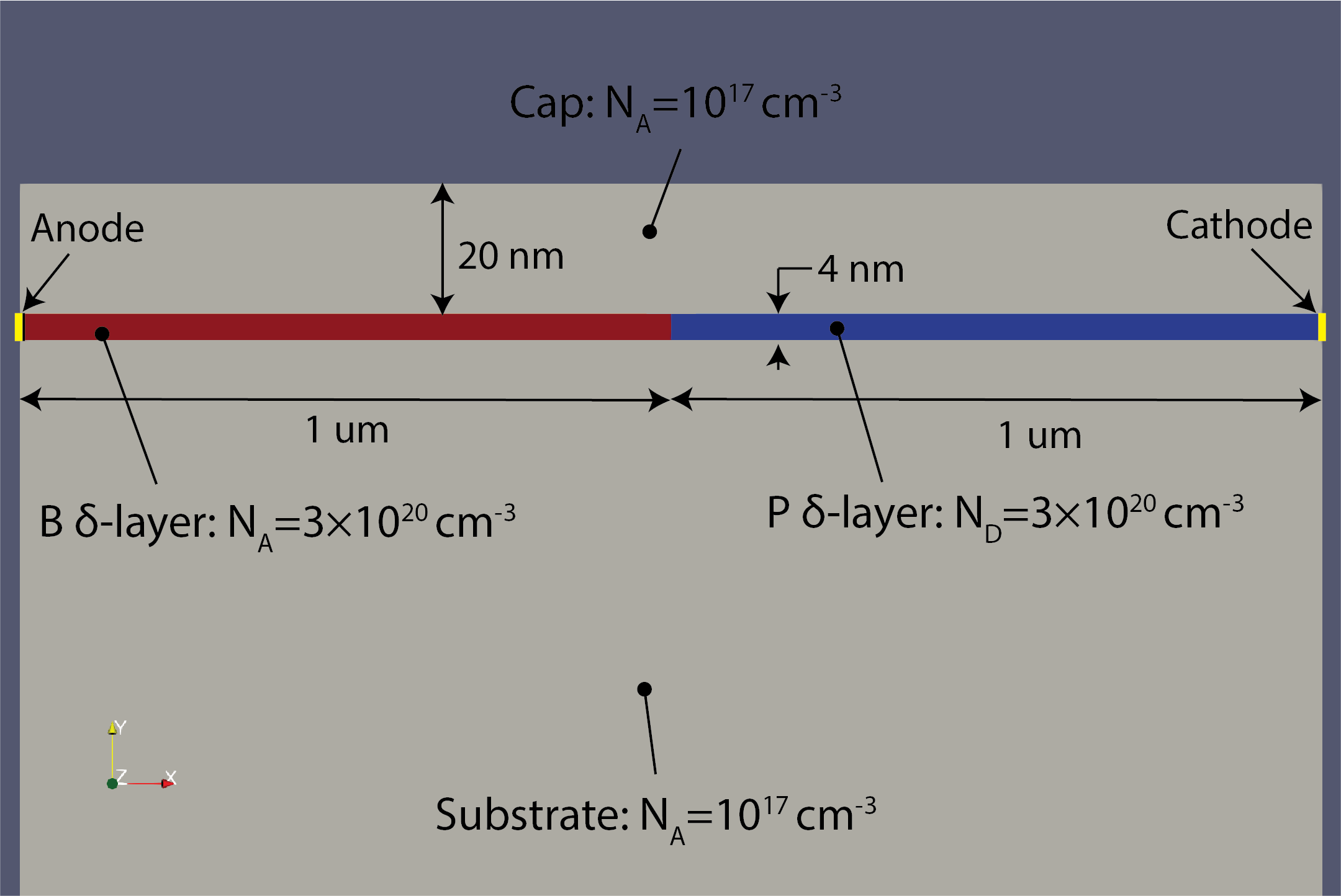}
  \caption{Schematic of the APAM pn junction device used in our simulations. It consists of a $\delta$-layer-based pn junction. The $\delta$-layers make direct Ohmic contact with the cathode and anode contacts as shown in the figure.}\label{fig:2D device}
\end{figure}

In our simulations, we employ the 2D structure shown in Fig.~\ref{fig:2D device}, which consists of a p-type $\delta$-layer with boron dopants and a n-type $\delta$-layer with phosphorous dopants embedded in silicon, known as APAM pn junction. The voltage is applied to the anode contact, while the cathode contact is grounded. The $\delta$-layers are approximated by highly-doped layers with a thicknesses of 4~nm. We chose a doping density of $3\times 10^{20}$~cm$^{-3}$ in the $\delta$-layers, and an acceptor density of N$_{A}=1\times 10^{17}$~cm$^{-3}$ in the Si substrate and cap. With the $\delta$-layer thickness being defined as $4$~nm and the doping density in the $\delta$-layers as $3\times 10^{20}$~cm$^{-3}$, this approximately results in the measured sheet doping density in the $\delta$-layers of $1.2\times 10^{14}$~cm$^{-2}$ in Ref.~\onlinecite{Skeren:2020}. Additionally, we chose a doping density for the substrate and cap that is several orders of magnitude lower than that in the $\delta$-layers, as is often the case in fabricated APAM devices.

As illustrated in Fig.~\ref{fig:2D device}, only the $\delta$-layers make direct Ohmic contact with the metallic anode and cathode, preventing unintended leakage paths through the substrate and cap. From the experimental observations in Ref.~\onlinecite{Skeren:2020} (see Fig.~\ref{fig:Measured I vs V curve}), electrons and holes are frozen-out in the enclosing substrate and cap regions at  temperatures below 20~K, and thus, the carrier transport mainly occurs across the $\delta$-layer-based pn junction; above 20-34~K, a significant number of dopants in the surrounding substrate become ionized, leading to an increase in current leakage through the substrate and cap. Since the primary aim of this work is to investigate the transport mechanisms through the $\delta$-layer-based pn junction, we chose to place the Ohmic contacts directly at the end of the $\delta$-layers. In other words, we assume that the contacts are isolated from the substrate and cap to avoid the undesired leakage current through the enclosing silicon (substrate and cap). 
However, to study the effect of current leakage on the I-V behavior, we have also included a qualitatively  analysis of leakage current through the substrate and cap in Sec.~\ref{sec: Current leakage} at 30~K.

\section{Transport Mechanism}\label{sec: transport mechanism}

\subsection{Band-to-band tunneling (BTBT)} \label{sec: BTBT}

\begin{figure*}[]
  \centering
  \includegraphics[width=0.9\linewidth]{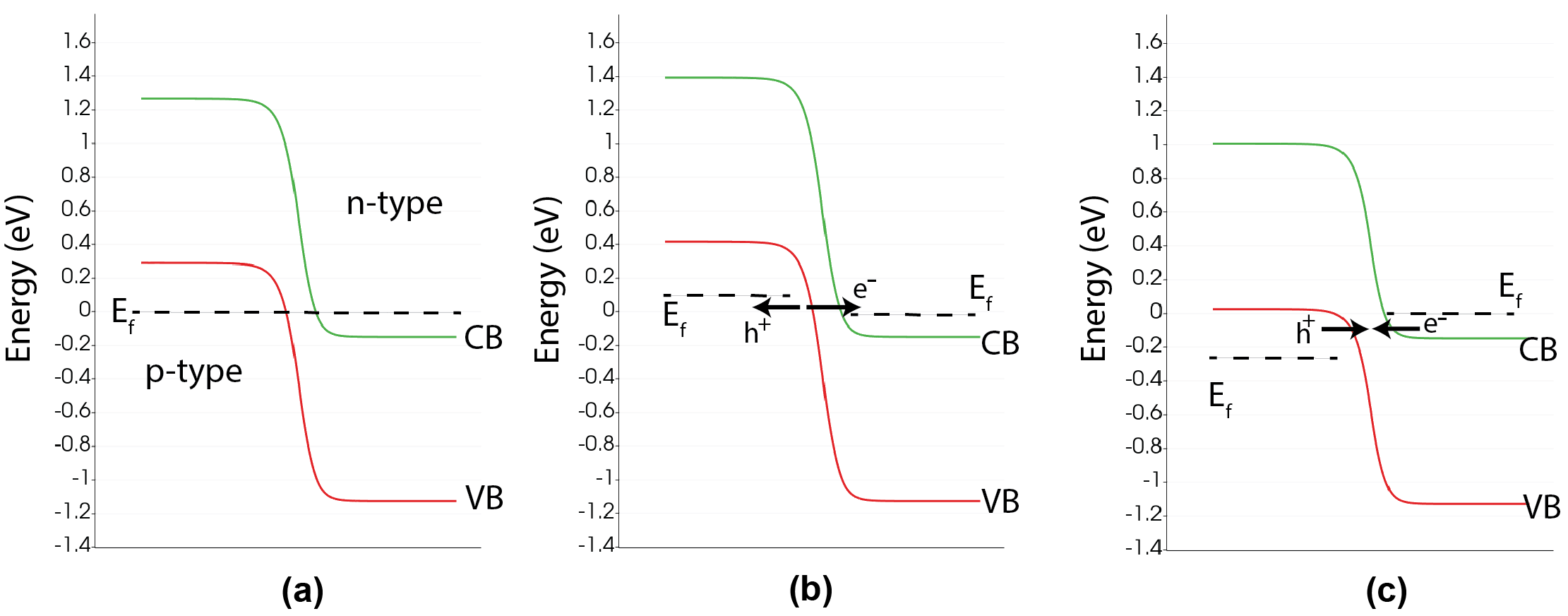}
  \caption{Conduction and valence band edges across the junction of an APAM pn device: \textbf{(a)} no tunneling in equilibrium condition; \textbf{(b)} tunneling generation in reverse bias; \textbf{(c)} tunneling recombination in forward bias.}
  \label{fig:BandStructureforB2Bt}
\end{figure*}

A typical band edge profile of an APAM pn junction under different bias conditions is shown in Fig.~\ref{fig:BandStructureforB2Bt}. Panels (\textbf{a}, \textbf{b}, \textbf{c}) represent the band edges across the junction under equilibrium condition, as well as under reverse and forward biased conditions, respectively. It is seen that the valence band of the p-type $\delta$-layer region overlaps with the conduction band of the n-type $\delta$-layer region, enabling the BTBT when a voltage is applied to the junction. In reverse bias, when a negative voltage is applied to the anode contact while the cathode contact is grounded (Fig.~\ref{fig:BandStructureforB2Bt}~\textbf{(b)}), the Fermi level of the anode contact increases in energy. This raises the valence and conduction bands of the p-type $\delta$-layer to higher energies, enabling the electron-hole generation, i.e., tunneling of valence electrons from the p-type $\delta$-layer to unoccupied conduction states in the n-type $\delta$-layer. Therefore, with increasing reverse voltages, the BTBT current will keep increasing thanks to the increased overlap. From Fig.~\ref{fig:Measured I vs V curve}, we observe that the current increases monotonically with negative voltages, a clear indication of BTBT tunneling (a.k.a. Zener tunneling) and in accordance with what was just described. 

On the contrary, in forward bias, when a positive voltage is applied to the anode contact (Fig.~\ref{fig:BandStructureforB2Bt}~\textbf{c}), the Fermi level of the anode decreases in energy. This causes the valence and conduction bands of the p-type $\delta$-layer to shift to lower energies, enabling the electron-hole recombination, i.e. electrons from the conduction band in the n-type $\delta$-layer tunnel into unoccupied valence states in the p-type $\delta$-layer. Quantum mechanically, the BTBT current is proportional to the product of the electron-hole wavefunction overlap and the difference in the Fermi-Dirac distributions. As the forward bias increases, the wavefunction overlap decreases, while the Fermi-Dirac distribution difference increases due to the rising energy difference between Fermi levels. The tunneling current is determined by the competing effect of these two factors. Initially, the increasing of Fermi-Dirac distribution difference dominates, causing the tunneling current to rise. But, at higher voltages, the reduction of wavefunction overlap becomes more significant, leading to a decline in current. Eventually, the wavefunction overlap approaches zero when the conduction and valence bands misalign, causing the BTBT to vanish. This BTBT mechanism leads to the well-known negative differential conductance (NDC) response in a tunneling diode under forward bias.

As described above, BTBT is a quantum mechanical process. In principle, a proper modeling of BTBT requires quantum mechanical evaluation, which involves solving self-consistently the Schr\"{o}dinger and Poisson equations. This type of quantum calculation is limited by a high computational burden and is not compatible with a PDE(partial differential equation)-based semi-classical TCAD code. Therefore, researchers over the past several decades have developed simplified, field-dependent BTBT models that are fully compatible with TCAD code. In Charon, the BTBT model in Eq.~\ref{eq:modified-b2bt} is implemented and can be used to approximate BTBT in pn junctions. The $A$ and $B$ parameters in the BTBT model (see Eq.~\ref{eq:modified-b2bt}) are related to material properties, such as effective masses and band gap\cite{Kao:2012,Kane:1960,Vandenberghe:2010}. These material properties can be significantly altered by the high doping densities and strong confinement of dopants in $\delta$-layers, causing the values of these parameters to differ substantially from those of bulk silicon. In the following, we evaluate how these parameters differ from bulk ones by fitting them with experimental data from Ref.~\onlinecite{Skeren:2020} under forward-bias conditions and examine the key different features between simulations and measurements.

\begin{figure}[ht]
  \centering
  \includegraphics[width=0.95\linewidth]{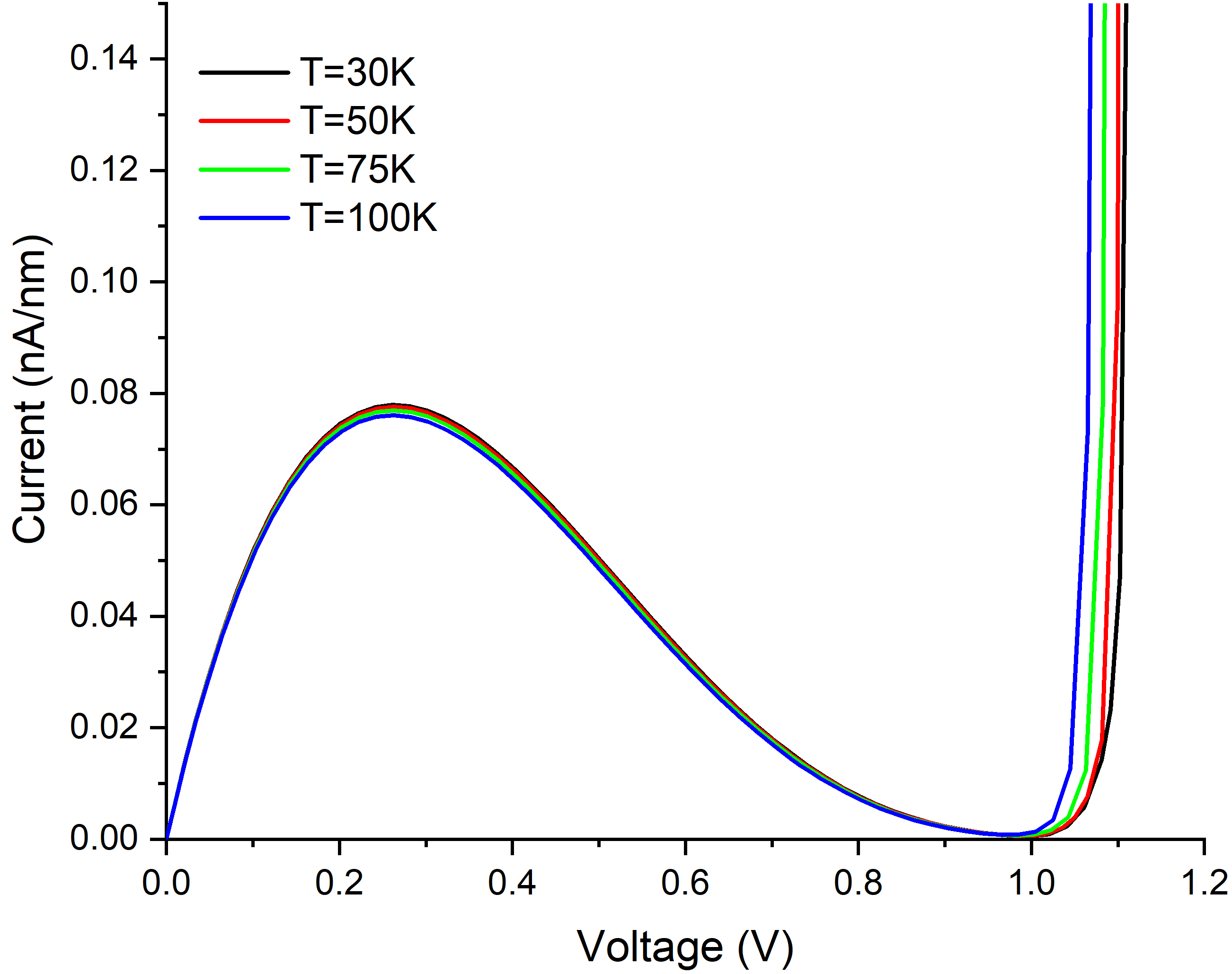}
  \caption{Simulated current-voltage (I-V) response of APM pn device at several low temperatures under forward bias. $A=1.0\times 10^{19}$ ~cm$^{-3}$s$^{-1}$, $B=3\times 10^{7}$~Vcm$^{-1}$, $\gamma$=2 and $\beta$=1. The $A$ and $B$ values correspond to bulk silicon.}
  \label{fig:forward I-V response: T study}
\end{figure}

As seen in Fig.~\ref{fig:Measured I vs V curve}, the measurements in Ref.~\onlinecite{Skeren:2020} were performed at very low temperatures as low as 1.7 K. We would ideally perform TCAD simulations at the same temperatures. However, TCAD simulations often fail to converge at low temperatures due to numerical issues arising from certain temperature-dependent material models, discretization or poorly conditioned solver matrices. Hence, we first study the temperature effect on the forward I-V response by simulating the APAM pn junction at various low temperatures and pushing the temperature to as low as the TCAD simulation can converge. For this study, we used the BTBT model with $A$ and $B$ values corresponding to bulk silicon. Although the BTBT model has no explicit temperature dependence, several material models contain temperature dependence such as mobility, band gap, and intrinsic concentration. The key material parameters for $\delta$-layers at the various temperatures are listed in Table~\ref{tab:Arora mobilities, BG and Electron affinity}. The simulated forward I-V curves are plotted in Fig.~\ref{fig:forward I-V response: T study}, where a NDC region is clearly observed, along with a valley voltage that agrees with the value expected by semi-classical models for a heavily doped silicon pn diode, which is nearly equal to the band-gap voltage. The NDC region is determined by BTBT and shows no temperature dependence as expected from the model. On the other hand, in the above-threshold region, i.e., for voltages above 1~V, the normal diode behavior shows a temperature dependence primarily due to the temperature variation of the band gap: the treshold voltage (a.k.a. valley voltage) decreases as the band gap increases with temperature. We also note that the temperature dependence in the TCAD simulations becomes weak for temperatures below 50~K as the result of very small changes in the material properties, such as band gap and mobilities, in highly doped systems within this low-temperature range. For example, the difference of band gap between 4-30~K is less than 0.1\%. Ref.~\onlinecite{Weir:1994} reported a little difference in hole mobilities for the range of temperatures between 4 and 77~K in very highly B doped silicon (see Table~\ref{tab:Arora mobilities, BG and Electron affinity}). Similar observation has been also reported for electrons in highly P doped silicon \cite{Goh:2006,Goh:2009,Ma:2012}. Since the TCAD simulation results (Fig.~\ref{fig:forward I-V response: T study}) show a negligible difference for temperatures below 50~K, and our TCAD simulations can converge  at the temperature as low as 30~K, all the following simulations and comparison with experiments have been done at 30~K, unless otherwise noted, while the mobilities were approximately set to the values indicated in Table~\ref{tab:Arora mobilities, BG and Electron affinity}. We emphasize that, under these conditions, the TCAD simulations resemble the conditions for the experimental measurements at the low temperatures of 1.7-10~K in Ref.~\onlinecite{Skeren:2020}.

\begin{figure}[ht]
  \centering
  \includegraphics[width=\linewidth]{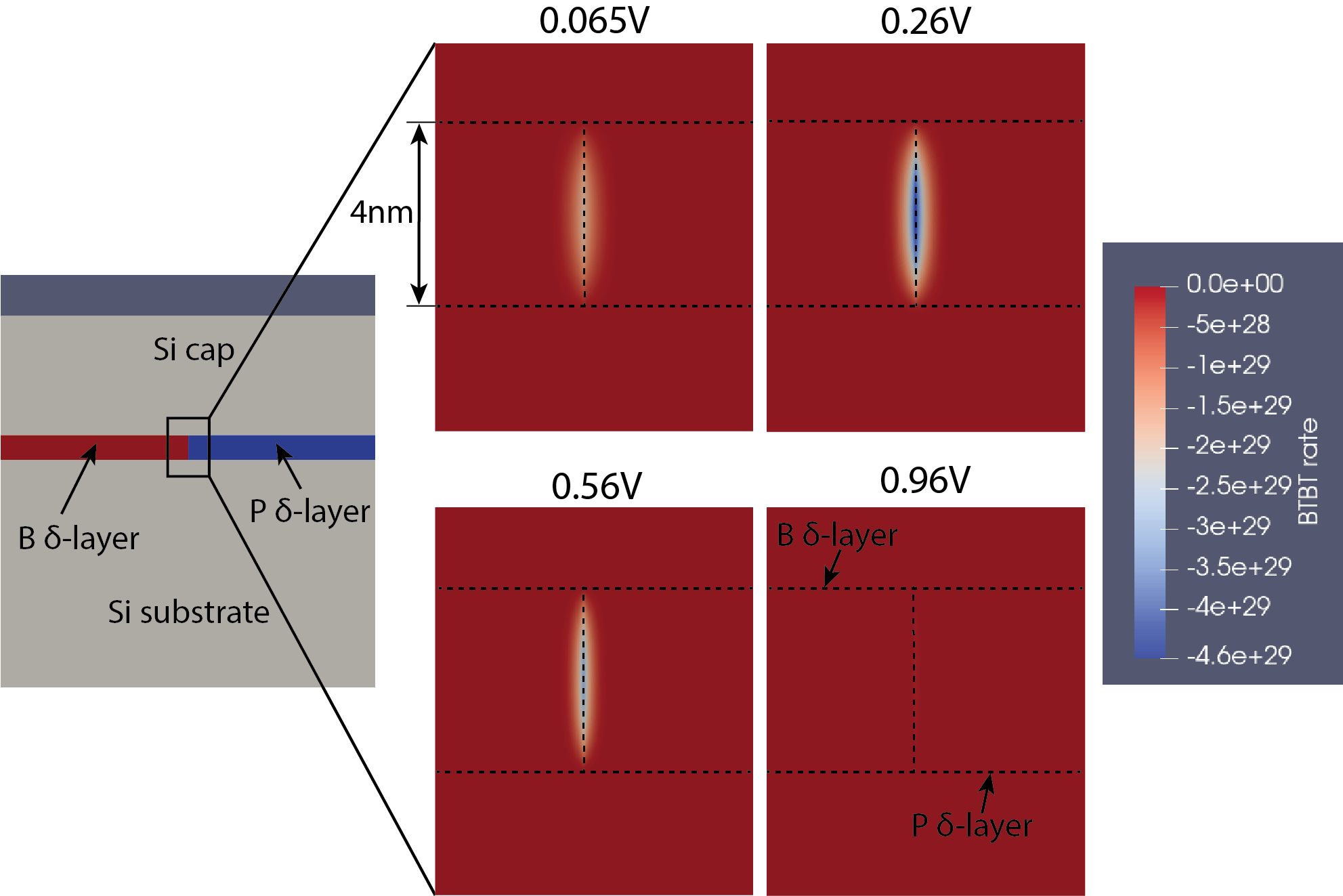}
  \caption{BTBT generation rate spatial profiles corresponding to several voltages in Fig.~\ref{fig:forward I-V response: T study}. The BTBT rate is in units of cm$^{-3}$s$^{-1}$. }
  \label{fig:BTBT_gen_rate}
\end{figure}

To further confirm that the NDC behavior in Fig.~\ref{fig:forward I-V response: T study} is due to BTBT, we plot the spatial profile of the BTBT generation rate near the pn junction for various voltages in Fig.~\ref{fig:BTBT_gen_rate}. The BTBT rate initially increases with increasing voltages, reaching its maximum magnitude at 0.26~V, which corresponds to the peak in the IV curve shown in Fig.~\ref{fig:forward I-V response: T study}. Afterward, the BTBT rate decreases, which corresponds to the current decrease observed in Fig.~\ref{fig:forward I-V response: T study}, and eventually vanishes above the valley voltage. The reason for the non-monotonic change of the BTBT generation rate in the forward-biased regime is due to the competing contributions from the field-dependent terms in Eq.~\ref{eq:modified-b2bt}. The first field-dependent term enforces zero BTBT under equilibrium conditions and its value increases as electric field decreases, while the other two field-dependent terms decrease with decreasing field. In the forward-biased regime, the junction's electric field keeps decreasing with increasing voltage. Thus, initially, the BTBT rate rises due to the dominance of the first field-dependent term but eventually declines as the other two field-dependent terms becomes more significant. The observed change of the BTBT rate with increasing forward voltage aligns with the expected behavior in highly doped pn junction, suggesting that our local, field-dependent BTBT model effectively captures the qualitative aspects of BTBT in these systems. However, we also indicate that this BTBT model does exhibit some overestimation because it does not account for electron-hole wavefunction overlap.

\begin{figure}[ht]
  \centering
  \includegraphics[width=0.95\linewidth]{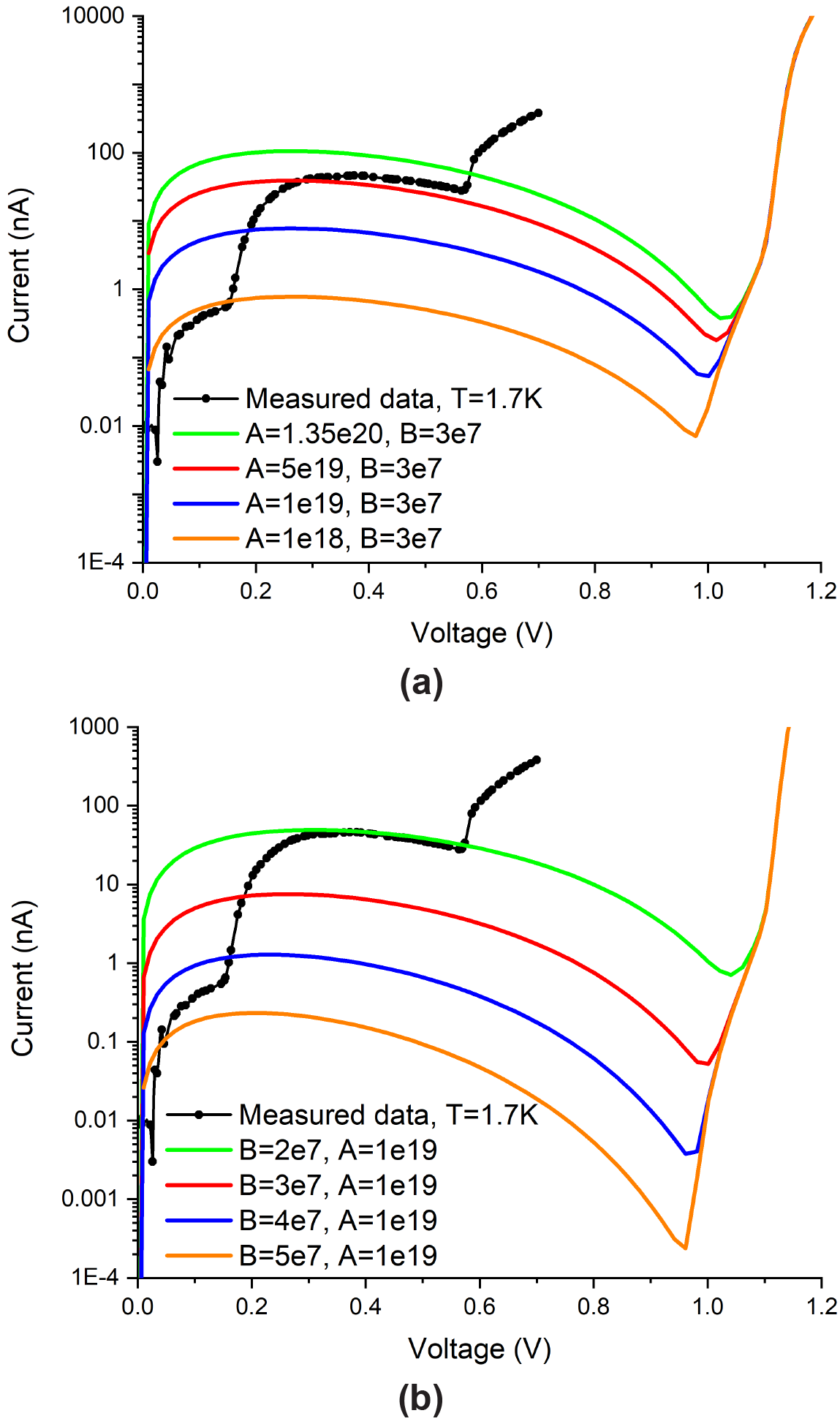}
  \caption{Simulated forward current-voltage (I-V) response. \textbf{(a)} For different values of the parameter $A$. \textbf{(b)} For different values of the parameter $B$. $\beta=1$, and $A$ and $B$ parameters are in units of cm$^{-3}$s$^{-1}$ and Vcm$^{-1}$, respectively. For comparison purposes, we multiplied the simulated currents in units of A/cm by the width of the $\delta$-layer of 100 nm, as given in Ref.~\onlinecite{Skeren:2020}.}
  \label{fig:foward I-V response: adjusting parameter}
\end{figure}

Fig.~\ref{fig:foward I-V response: adjusting parameter} includes an analysis of the effect of $A$ and $B$ parameters on the I-V response within the NDC region. From panel \textbf{(a)}, currents in the NDC region increase proportionally with increasing $A$ values, while the NDC peak-current voltages and valley voltages are not affected by different $A$ values. From panel \textbf{(b)}, it can be observed that currents in the NDC region increase as the $B$ value decreases, and a smaller $B$ value shifts the peak-current voltage and the valley voltage slightly to the right. 
Comparing the simulated forward I-V curves to the measured data in Fig.~\ref{fig:foward I-V response: adjusting parameter}, we observe two major differences: (i) the valley voltage in the measured data is much smaller than the theoretical value of about 1~V; (ii) the current suppression between 0 and 0.15~V in the experimental data is not seen in the simulated curve. 
The question that arises is: what possible physical mechanisms could cause these differences? To address these unexplained behavior, we investigate three potential causes in the next subsections: (i) significant band-gap narrowing resulting from the modification of the band structure induced by the $\delta$-layers (Sect.~\ref{subsection:BGN}); (ii) non-ideal contact such as Schottky contact at the B $\delta$-layer (Sect.~\ref{sec: effect of schottky contact}); and (iii) band quantization due to the strong confinement of the dopants in the $\delta$-layers (Sect.~\ref{sec: Effect of band quantization}). 

\subsection{Effect of band gap narrowing}\label{subsection:BGN}

Band gap narrowing (BGN) is the reduction of the energy difference between the conduction and valence band edges in a semiconductor material caused by the effect of increasing temperature and/or doping densities. In TCAD simulations, various models, such as the Slotboom model \cite{Slotboom:1977}, have been employed to incorporate this effect. These models are known to work well for lightly to moderately doped semiconductors. However, they fail to accurately represent the impact of highly doped materials, such as in $\delta$-layers. Additionally, they only consider the total doping density, without distinguishing between n-type or p-type dopants.

\begin{figure}[ht]
  \centering
  \includegraphics[width=0.95\linewidth]{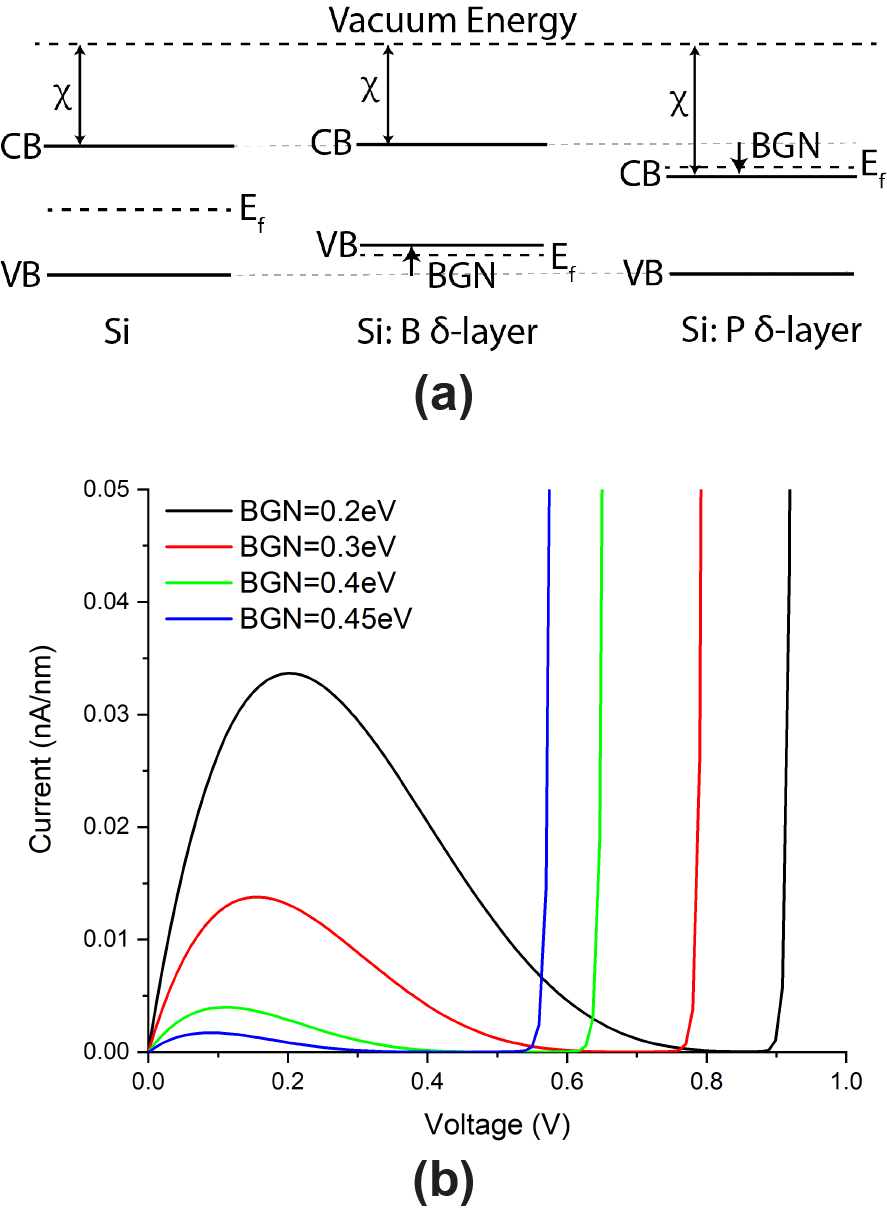}
  \caption{Effect of the band gap narrowing (BGN). \textbf{(a)} Schematic band diagram for the proposed BGN model within the p-type and n-type $\delta$-layer regions; \textbf{(b)} Study of the effect of the corresponding BGN for the forward-bias response. BTBT parameters used in the simulations are: $A=1\times 10^{19}$~cm$^{-3}$s$^{-1}$, $B=3\times 10^{7}$~V/cm, and $\beta=1.0$. }
  \label{fig:BGN}
\end{figure}

It is known that the high doping densities in $\delta$-layers significantly modify the electronic band structure, as observed in experiments \cite{Miwa:2013,Miwa:2014,Mazzola:2018,Holt:2020,Mazzola:2020} and quantum-mechanical simulations \cite{Mamaluy:2021} for n-type $\delta$-layers, and in DFT simulations for p-type $\delta$-layers \cite{Campbell:2023}. To overcome the limitation of standard BGN models and account for the significant effect of BGN in $\delta$-layers, we explore the potential impact of BGN in $\delta$-layers using the model shown in Fig.~\ref{fig:BGN}\textbf{(a)}. In contrast to standard models, the band gap in both $\delta$-layer regions are reduced with the doping density, but the electron affinity is only modified for the n-type $\delta$-layer region. Fig.~\ref{fig:BGN}\textbf{(b)} shows the simulated I-V responses under forward voltages for different BGN values, with the same magnitude of BGN applied to both $\delta$-layer regions. From these results, we observe that, as the BGN value increases, the valley voltage decreases due to the band-gap reduction in the $\delta$-layers. This is accompanied by a reduction in current within the NDC region and a shift of the peak voltage to lower values. This occurs because the magnitude of the electric field in the pn junction decreases, leading to a reduction of the BTBT rate. Therefore, the BGN effect, induced by $\delta$-layers can plausibly explain the small valley voltage observed in the experimental data for APAM pn junctions in Ref.~\onlinecite{Skeren:2020}. From these results, a BGN of 0.45~eV leads to a valley voltage value close to that observed in Fig.~\ref{fig:Measured I vs V curve}. This magnitude of BGN is also in very good agreement with ARPES measurements from Refs.~\onlinecite{Miwa:2013,Mazzola:2018}.

In previous works\cite{Chynoweth:1961,Rivas:2003}, the observed lower valley voltage than the theoretically expected value in Si Esaki diodes\cite{Skeren:2020,Chynoweth:1961} was explained by the existence of density of states inside the band gap. These states facilitate the direct coherent tunneling, from gap state to gap state, leading to what they call "excess current". This current determines the valley voltage, and because it is driven by direct tunneling, it is temperature independent\cite{Rivas:2003}. In highly doped system, e.g., $\delta$-layer systems, those states become continuous in space, thus lowering/raising effectively the conduction/valence band edge and reducing the band gap. This observation substantiates our semi-classical approach of using the BGN effect to simulate the measured low valley voltage.

\subsection{Effect of Schottky contact}\label{sec: effect of schottky contact}

One of the hypotheses for the current suppression at the low voltage onset is the existence of a non-Ohmic contact. Thus, we next investigate the potential effect of a Schottky contact on the p-type side of the bipolar device, since the Schottky barrier is likely to be higher on that side. The existence of a Schottky barrier for holes at the junction between the metallic contact and the p-type $\delta$-layer is due to the work function difference between them, being lower at the metal than at the p-type semiconductor. When applying a negative bias to the Schottky contact, the barrier for holes in the p-type semiconductor decreases, thus increasing the thermionic emission current; otherwise, when applying a positive bias to the Schottky contact, the barrier height for holes increases, thus reducing the thermionic emission current. However, for very highly doped semiconductors, tunneling plays an important role in the latter case, allowing holes to pass through the Schottky barrier. Higher doping density narrows the barrier width (i.e., the depletion region width), thereby increasing tunneling. Counterintuitively, with increasing positive bias, the barrier height increases, but the barrier width decreases and the electric field increases, thereby enhancing the current through tunneling. We also note that, at low temperature, the effect of temperature on the hole tunneling is weak.

\begin{figure}[ht]
  \centering
  \includegraphics[width=0.95\linewidth]{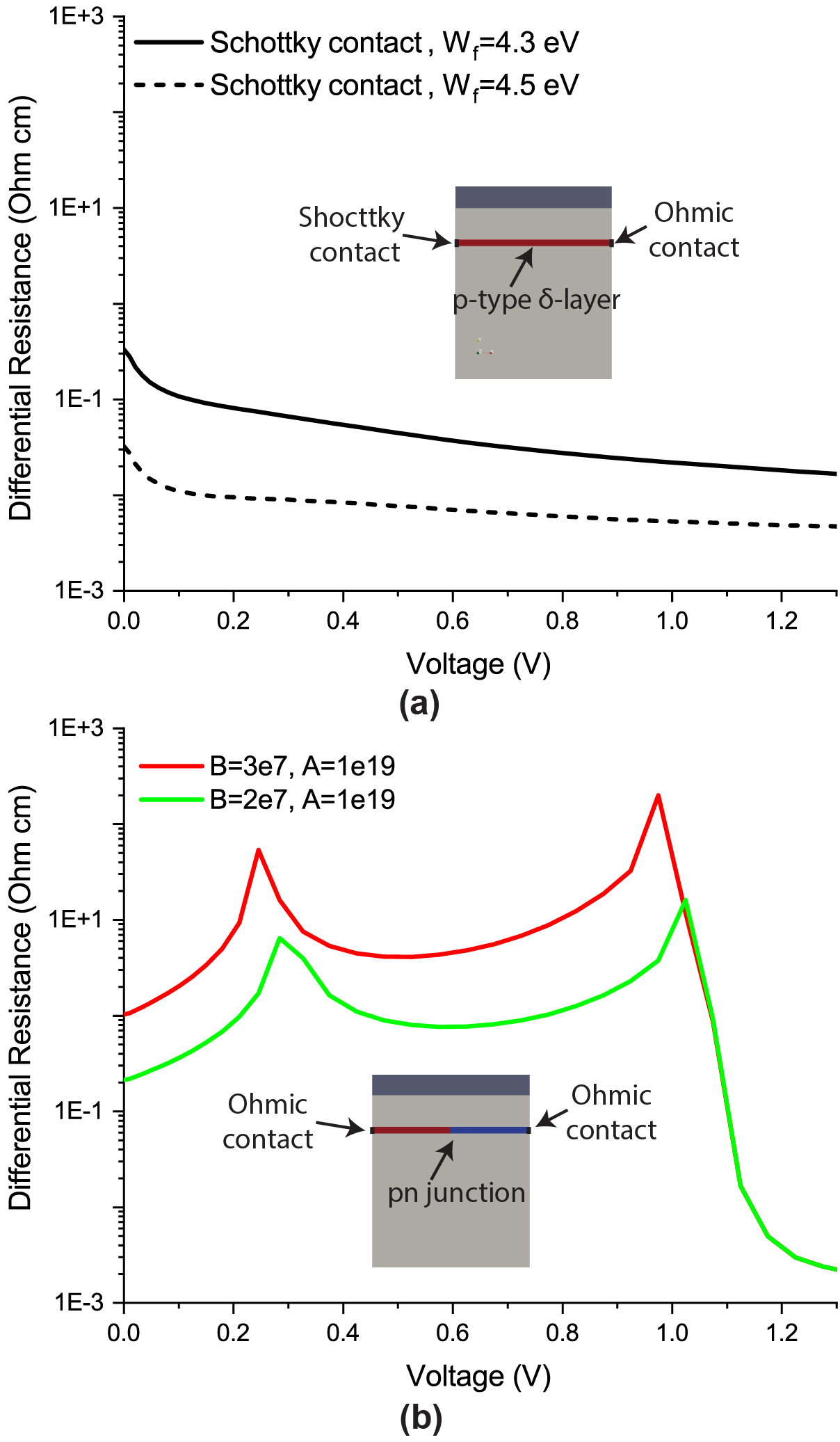}
  \caption{Study of the Schottky contact effect. \textbf{(a)} Absolute value of the differential resistance of the Schottky contact between a very highly p-type doped $\delta$-layer and a metal, for two different work function ($W_f$) values. \textbf{(b)} Absolute value of the differential resistance of the APAM pn junction. $A$ and $B$ parameters are in unit of cm$^{-3}$s$^{-1}$ and Vcm$^{-1}$, respectively. The chosen metal work function values correspond to that of Al ($W_f=4.3$~eV), which was used as the contact material for the bipolar device in Ref.~\onlinecite{Skeren:2020}, and to a metal with a higher work function. }
  \label{fig:Study of the Schottky contact effect}
\end{figure}

The first question to be addressed is whether the Schottky barrier is transparent to holes in very highly doped p-type $\delta$-layers (e.g. $3 \times 10^{20}$~cm$^{-3}$), i.e., whether holes can tunnel through the barrier or are blocked. Fig.~\ref{fig:Study of the Schottky contact effect}~\textbf{(a)} shows the Schottky contact resistance computed for a p-type $\delta$-layer and a metal, for two different work function values ($W_f=4.3,4.5$~eV). The chosen work function values correspond to that of Al ($W_f=4.3$~eV), which was used as the contact material for the bipolar device in Ref.~\onlinecite{Skeren:2020}, and to a metal with a higher work function. The Schottky barrier height for a p-type semiconductor-metal is defined as $\Phi_p=E_g-(W_f-\chi_s)+V_{app}$, where $E_G$ is the band-gap, $W_f$ is the metal work function, $\chi_S$ is the electron affinity in the semiconductor and $V_{app}$ is the voltage applied to the Schottky contact. Therefore, the barrier height is approximately of $0.9(0.7)+V_{app}$~eV for a metal work function of 4.3(4.5)~eV. We first note that, for very highly doped $\delta$-layers, the resistance in the $\delta$-layers is negligible, so the resistance in \textbf{(a)} is mainly due to the Schottky contact contribution. From the results in the figure, we observe that the resistance decreases as the metal work function increases. These results also indicate that the overall contact resistance is low, with the Schottky barrier being nearly transparent to holes despite its high value. Additionally, we observe no evidence of an abrupt transition in resistance within the voltage range of 0 to 0.2~eV, which would be necessary to explain the transition seen in Figure~\ref{fig:Measured I vs V curve}. Finally, we note that, as the bias increases, the Schottky resistance decreases further, enhancing the transparency of the barrier and facilitating hole tunneling.

For comparison, Fig.~\ref{fig:Study of the Schottky contact effect}~\textbf{(b)} shows the resistance of the pn junction from Fig.~\ref{fig:2D device} with Ohmic contact on both sides. Note that the pn junction resistance exhibits two peaks corresponding to the peak current and valley voltage locations. Comparing the magnitude of the resistance in both cases (\textbf{(a)} and \textbf{(b)}) reveals that the pn junction resistance is generally at least one order of magnitude higher than the Schottky contact resistance in the voltage range of interest. Together with the absence of evidence for an abrupt transition in the Schottky contact resistance within the voltage range of 0 to 0.2~V, we can conclude that Schottky contact is unlikely the reason for current suppression in the forward low-bias region.

\subsection{Effect of band quantization}\label{sec: Effect of band quantization}

\begin{figure}[ht]
  \centering
  \includegraphics[width=0.95\linewidth]{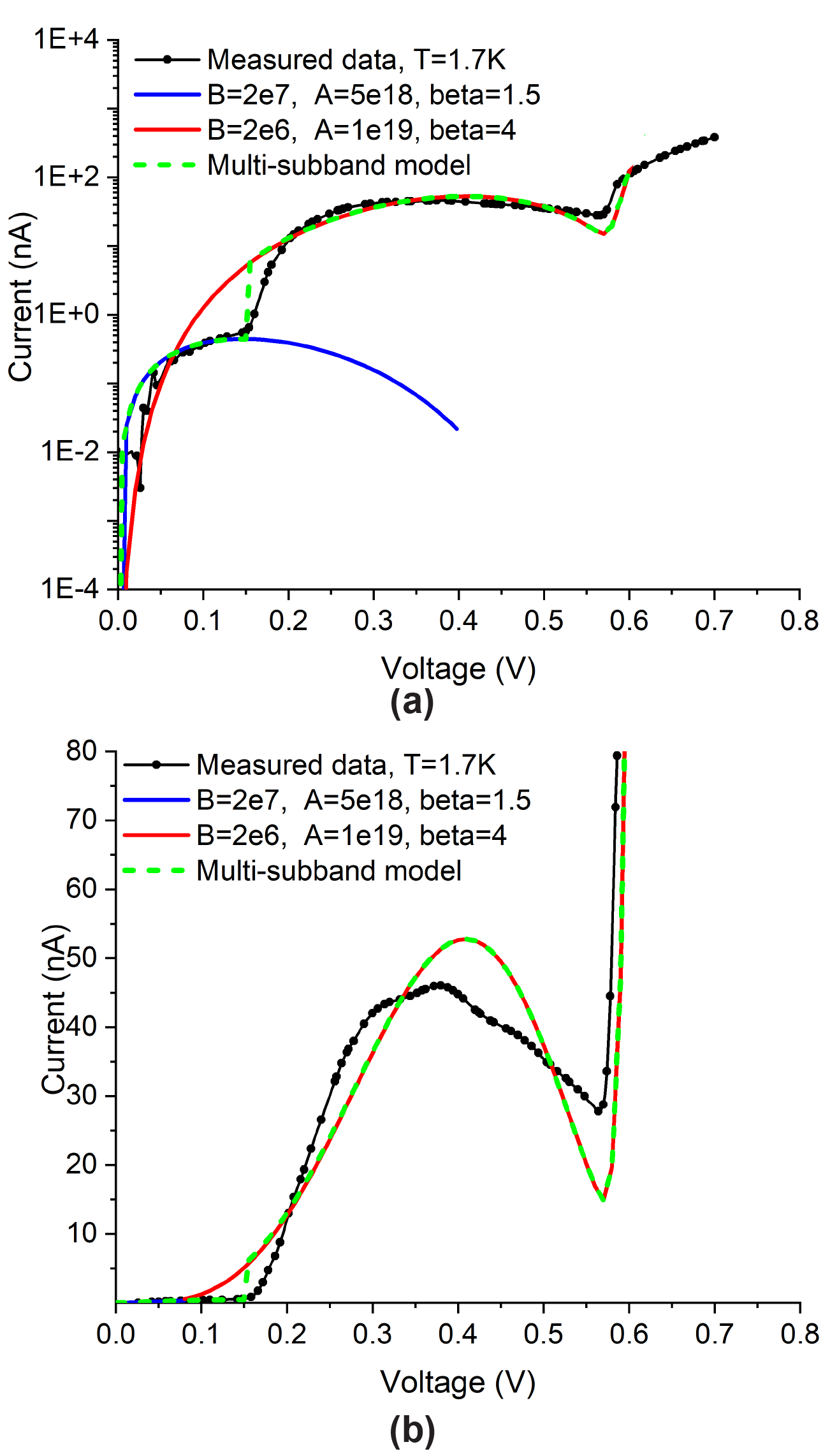}
  \caption{Comparison of the simulated I-V response with the measured data using two sets of fitted BTBT parameters and the BGN model presented in Section~\ref{subsection:BGN} with a BGN of 0.45~eV. \textbf{(a)} Semi-log scale, and \textbf{(b)} linear scale. $A$ and $B$ parameters are in units of cm$^{-3}$s$^{-1}$ and Vcm$^{-1}$, respectively. For comparison purposes, we have multiplied the simulated currents in units of A/cm by a width of the $\delta$-layer of 100~nm, as given in Ref.~\onlinecite{Skeren:2020}.}
  \label{fig:Comparison I vs V curve}
\end{figure}

Our simulations indicate that the main features in the measurements within the voltage range of 0-0.6~V can be accounted for by using different combinations of BTBT parameters.
As shown in Fig.~\ref{fig:Comparison I vs V curve}, using the BTBT parameter set, $A = 5.0\times 10^{18}$~cm$^{-3}$s$^{-1}$, $B = 2.0\times 10^7$~Vcm$^{-1}$ and $\beta=1.5$, our simulation can capture the lower elbow of the I-V curve in the 0.0-0.15~V region, while using the BTBT parameter set, $A = 1.0\times 10^{19}$~cm$^{-3}$s$^{-1}$, $B = 2.0\times 10^6$~Vcm$^{-1}$ and $\beta=4$, our simulation can capture reasonably well the higher elbow of the I-V curve in the 0.15-0.6~V region. The low valley voltage is well modeled with a BGN value of 0.45~eV as reported in the previous section. Thus, using these two sets of BTBT parameters and the BGN of 0.45~eV, we can qualitatively capture the two distinct forward I-V regions. This indicates that there may be two distinct BTBT regimes at work which can stem from band quantization and realignment of these quantized states.
This finding aligns very well with recent experiments\cite{Holt:2020, Mazzola:2020} and simulations\cite{Mamaluy:2021,Mendez:2022}, where the existence of quantized conduction bands in n-type $\delta$-layers and quantized valence bands in p-type $\delta$-layers were revealed. The band quantization was attributed to the strong confinement of dopants in one direction (a.k.a. size quantization). Additionally, in Refs.~\onlinecite{Mamaluy:2021,Mendez:2022}, it was found that the number of quasi-quantized states in n-type $\delta$-layers, as well as the splitting energy between them, are strongly dependent on both the $\delta$-layer thickness and the doping density. We also note that there is a discrepancy between our theoretical prediction and the measurements, specifically the low predicted valley current. This is likely associated with other mechanisms we have not taken into account here, such as elastic tunneling through traps.

\begin{figure*}[]
  \centering
  \includegraphics[width=0.95\linewidth]{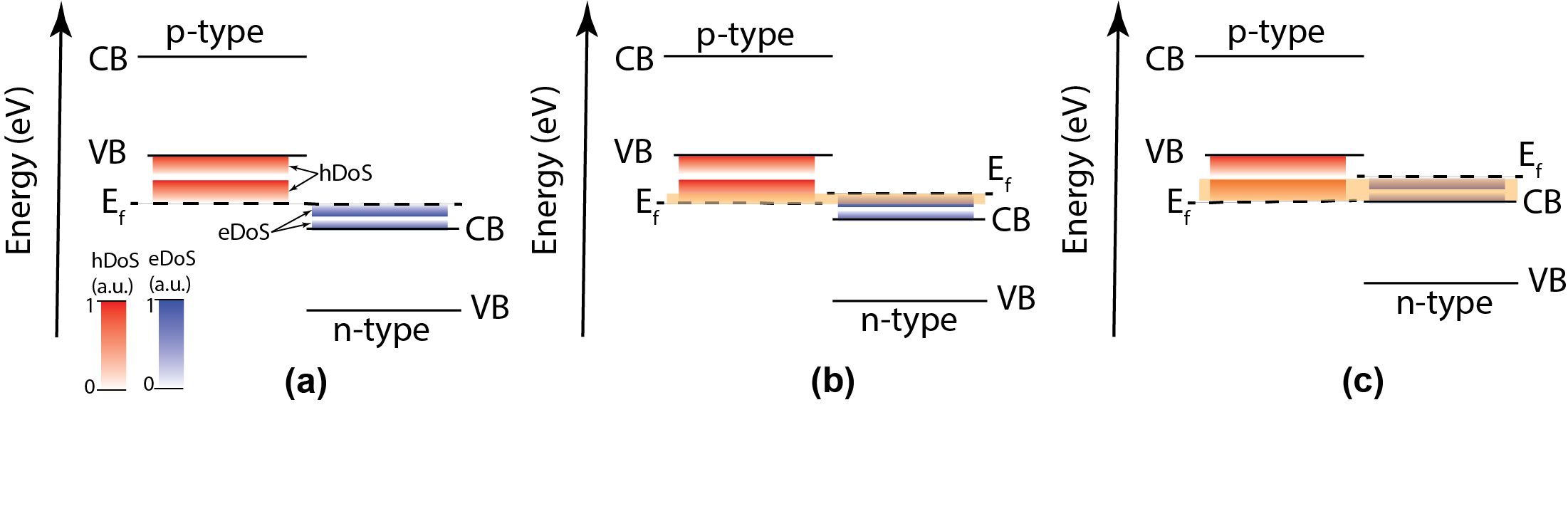}
  \caption{Quantization of the band structure in $\delta$-layers. The figure represents the band structure across the pn junction: \textbf{(a)} band structure in equilibrium; \textbf{(b)} after applying a small forward bias of V$_0$;  \textbf{(c)} after applying a higher forward bias V$_1$$>$V$_0$. hDoS (in red) represents the hole density of states, while eDoS (in blue) represents the electron density of states. The yellow highlight represents the overlap between hole states and electron states.}
  \label{fig:Band quantization}
\end{figure*}

To gain further understanding, a schematic representation of a possible combination of band quantization is shown in Fig.~\ref{fig:Band quantization}, assuming that only two quantized subbands exist in both conduction and valence band structures of the n-type and p-type $\delta$-layers, respectively. The two shaded areas in the valence and conduction bands represent the densities of occupied states for holes and electrons, respectively, in the two quantized sub-bands. As observed from the figure, initially, when a forward bias is applied, only two subbands overlap (see \textbf{(b)}). With a further increase in the bias, more subbands will be involved (see \textbf{(c)}), leading to a potential increase in the current. Thus, a very plausible explanation for the atypical behavior shown in region 2 and 3 in Fig.~\ref{fig:Measured I vs V curve} might be attributed to the quantization of the band structure and the realignment of these quantized states. It is possible that the first BTBT region (between approx. 0 and 0.15~V) occurs when the lowest conduction subband and highest valence subband are only aligned. As the applied voltage is increased, additional subbands align, enhancing BTBT and giving rise to the second BTBT regime (between approx. 0.15 and 0.6~V). This step-like behavior due to different band alignment has been also reported in germanium electron-hole bilayer tunnel FET\cite{Alper:2013}. Full quantum mechanical modeling of BTBT in an APAM pn junction in the future will clarify this hypothesis.

The BTBT in APAM pn junctions can be modeled using a multi-subband model based on the Hurkx’s BTBT model by taking
\begin{equation}
G_{btbt}=\sum_{i}^n G_{btbt}^i
\label{eq:multi-subband model}
\end{equation}
where $n$ is the total number of overlapping subbands between valence subbands in the p-type $\delta$-layer and conduction subbands in the n-type $\delta$-layer. $G_{btbt}$ follows the modified Hurkx's model in Eq.~\ref{eq:modified-b2bt}. In particular, for the APAM pn junction from Ref. \onlinecite{Skeren:2020}, two regimes are observed; thus, Eq.~\ref{eq:multi-subband model} takes the form $G_{btbt}=G_{btbt}^{I} + G_{btbt}^{II}$, where $G_{btbt}^{II}$ only exists when $V>0.15$~V. $G_{btbt}^I$ and $G_{btbt}^{II}$ have different set of parameters, and those parameters are not only related to the material properties, but also to the $\delta$-layer thickness. Fig.~\ref{fig:Comparison I vs V curve} includes the IV curve predicted using this multi-subband model.

\subsection{Current leakage}\label{sec: Current leakage}

In previous sections, we have assumed that the $\delta$-layers only form contacts with the anode/cathode, and not the substrate or cap. This is justified by the fact that at low temperature, the leakage currents across the Si substrate and cap are minimal due to the dopants in those areas being inactive at cryogenic temperatures. However, at higher temperatures, the dopants will become ionized, opening the possibility of significant leakage currents, which was speculated to occur in Ref.~\onlinecite{Skeren:2020} at temperatures above 20~K. In this section, we address the possibility of leakage and its impact on the simulated I-V response.

\begin{figure}[ht]
  \centering
  \includegraphics[width=0.95\linewidth]{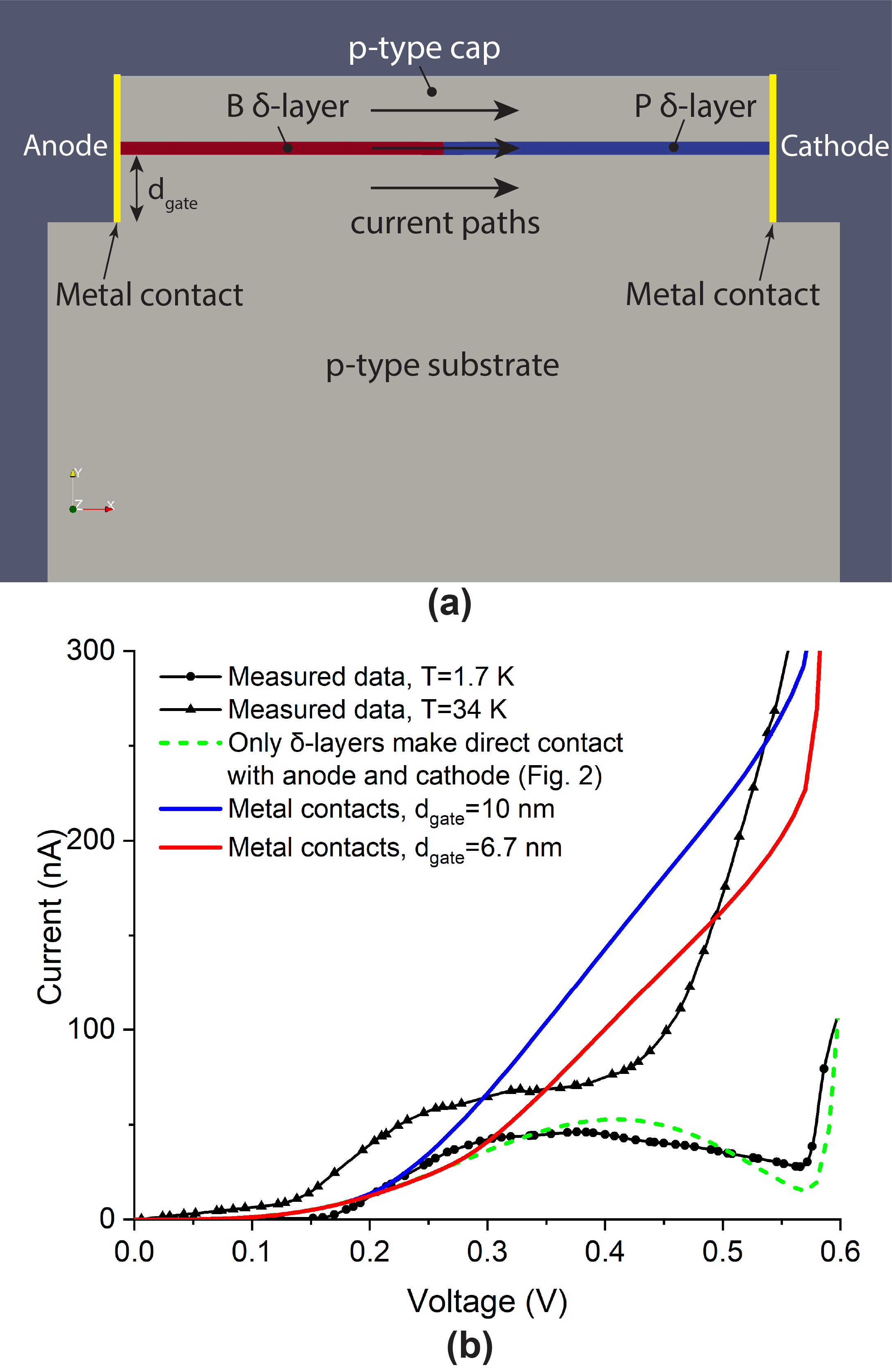}
  \caption{\textbf{(a)} APAM pn device with metal contacts that touch enclosing Si cap and substrate. \textbf{(b)} Study of the current leakage at approx. 30~K for the configurations presented in \textbf{(a)}. $T=30$~K, $A=1\times 10^{19}$~cm$^{-3}$s$^{-1}$, $B=2\times 10^{6}$~V/cm, $\beta=4.0$, and BGN=0.45~eV.}
  \label{fig:leakage study} 
\end{figure}

To simulate the leakage current through the substrate and cap regions, we extend the anode and cathode contacts to cover the cap region and part of the substrate, as shown in Fig.~\ref{fig:leakage study}~\textbf{(a)}. For this scenario, we consider two contact penetration depths into the substrate: $d_{gate}=6.7$,  $10$~nm. 
Fig.~\ref{fig:leakage study}~\textbf{(b)} includes the simulated I-V curves alongside the experimental data from Ref.~\onlinecite{Skeren:2020} at the temperature of 1.7~K, where leakage is negligible, and at the temperature of 34~K, where leakage starts to become non-negligible. We first note that the simulated I-V curve (dashed green) without leakage qualitatively reproduces the measured I-V at 1.7~K, as discussed in previous sections. In contrast, the simulated I-V curves (continuous blue and red), which account for leakage through the substrate and cap regions, deviate significantly from the leakage-free case. Specifically, the NDC behavior observed in the leakage-free I-V curve disappears when leakage is included, matching qualitatively the measured I-V characteristics at 34~K. Furthermore, the current, including leakage, increases significantly as the contact depth into the substrate increases. Therefore, leakage through the substrate and cap regions could indeed lead to the increased measured current and the disappearance of NDC at 34~K, as suggested in Ref.~\onlinecite{Skeren:2020}. Since the geometry of the actual contact in the measured device is unknown, it is not possible to quantify the leakage and hence we chose not to try to match simulation results with the data at 34~K.

From this study, we observe that the performance of APAM devices can be significantly impacted by leakage through the substrate/cap at elevated temperatures. Therefore, when designing APAM devices for elevated-temperature operation, it is crucial to minimize the current leakage, potentially achieved by using a Silicon on Insulator (SOI) substrates as proposed in Ref.~\cite{Skeren:2018}, where the thin device will minimize the potential leakage pathways.

\subsection{Trap-assisted tunneling}\label{TAT at room temperature}

In previous work\cite{Mazzola:2014}, it was shown that the contribution of electron-phonon scattering towards the $\delta$-layer resistance was minimal for temperatures below 100~K, with the electron-impurity scattering being the largest contribution. This finding is also sustained by the observation of little change in the electron and hole mobilities between 4-77~K \cite{Weir:1994}. Hence. the contribution from inelastic trap-assisted tunneling mediated by phonon interactions at temperatures below 100~K is expected to be negligible, although the contribution from elastic trap-assisted tunneling might be appreciable. However, at elevated temperatures (above 100~K), trap-assisted tunneling (TAT) due to electron-phonon interactions may become relevant, particularly at room temperature. In this section, we investigate how phonon-enabled TAT might affect the device response at room temperature.

\begin{figure}[ht]
  \centering
  \includegraphics[width=0.95\linewidth]{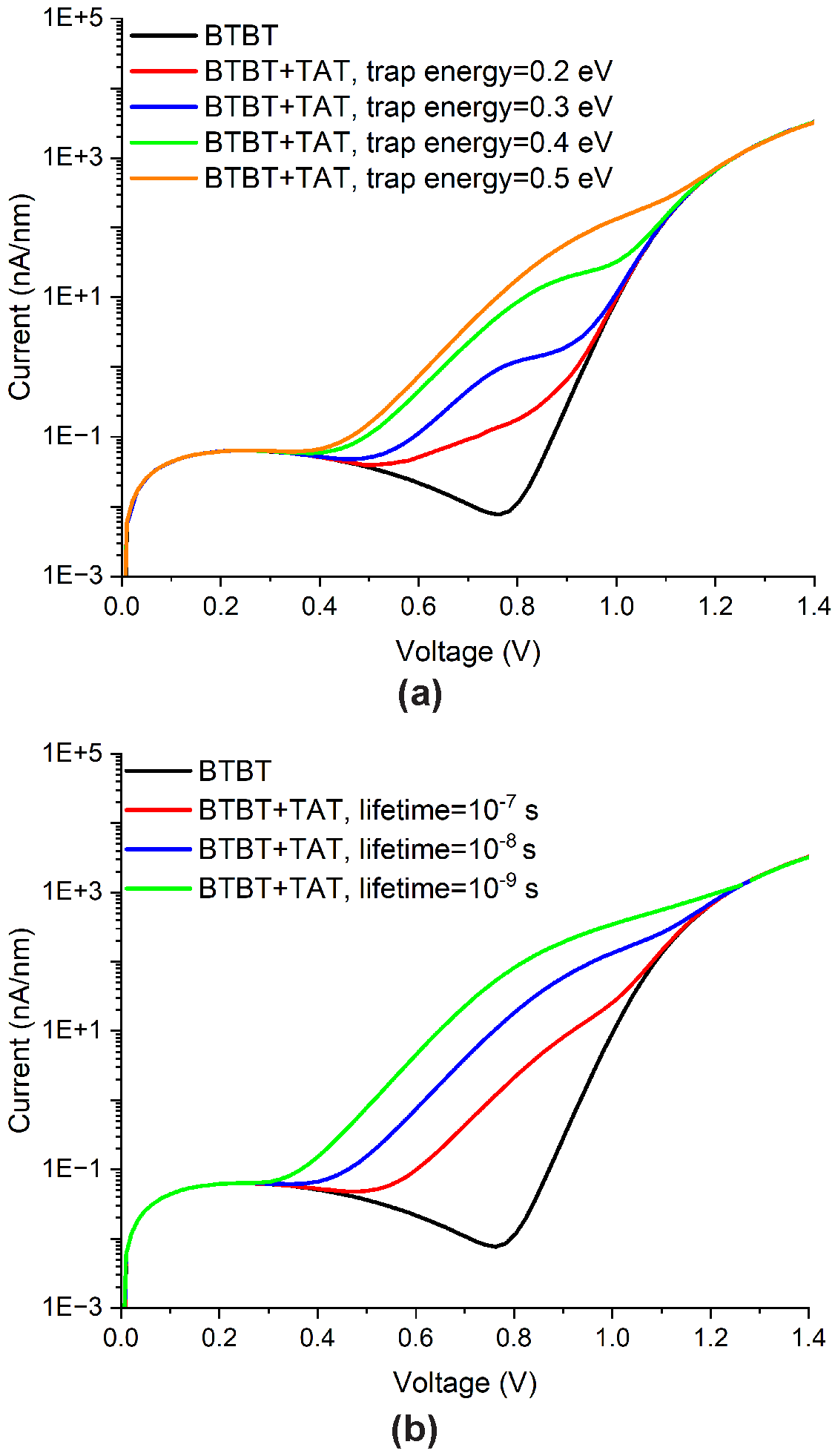}
  \caption{Study of trap assisted band tunneling (TAT) at room temperature in APAM pn junctions: \textbf{(a)} effect of trap energy levels, and \textbf{(b)} effect of lifetimes. BTBT parameters used in the simulations are: $A=1\times 10^{19}$~cm$^{-3}$s$^{-1}$, $B=3\times 10^{7}$~V/cm, and $\beta=1.0$. Default values for the parameters in the TAT model are: lifetimes $\tau_{n0}=\tau_{p0}=10^{-8}$~s, and trap energy level E$_t$=0.5~V. Details of the TAT parameters can be found in Appendix~A.}
  \label{fig:tat_study_fitted_parameters}
\end{figure}

For our simulations, we use the TAT model developed by Schenk in Ref.~\onlinecite{Schenk:1992}. A summary of the model implementation is included in Appendix~A. The following tunneling relevant parameters are kept constant in our simulations: Huang–Rhys factor of $3.5$ and the optical phonon energy is assumed to be $\hbar \omega =68$~meV. Fig.~\ref{fig:tat_study_fitted_parameters} depicts the simulated I-V forward responses including TAT near the junction for different trap energy levels in \textbf{(a)} and for different carrier lifetimes in \textbf{(b)}. We note that the trap energy in this model is with respect to the conduction band edge. Interestingly, the simulated I-V curves show in Fig.~\ref{fig:tat_study_fitted_parameters} three distinct regions: (i) currents at low voltages are determined by BTBT; (ii) currents at medium voltages are dominated by TAT, showing a strong dependence on TAT model parameters; (iii) currents at high voltages result from a resistor-like response. These results show that TAT rates decrease for shallower trap defects, i.e., traps with energy levels closer to the conduction band edge. This is consistent with the fact that deep-level traps are more efficient recombination centers. In addition, TAT rates decrease as the carrier lifetimes become larger. The lifetime is inversely related to the number of defects near the junction, i.e., the lower defect density, the higher lifetime of the carrier. 

The Poole-Frenkel effect, which describes how the effective depth of a trap is reduced due to the Coulomb interaction between a free carrier and the trap, is not taken into account in our TAT model. However, it is important to note that this effect is weaker than the tunneling effect at strong electric fields, whereas it might moderately enhance the emission probability at weak electric fields, and thus influence the net recombination rate\cite{Hurkx:1992}. Given that the built-in electric field across the pn junction in our device is very high, we expect that the Poole-Frenkel effect on the device response will be minimal at very low forward biases, and it might only slightly enhance the current in the voltage range where the TAT is significant. 

From the results of Fig.~\ref{fig:tat_study_fitted_parameters}, it is evident that the effect of TAT will decrease as the junction becomes more ideal. Conversely, it can become the dominant mechanism if there are a considerable number of defects near the junction. Therefore, from this simulation study, we predict that, as APAM pn devices become more mature to operate at room temperature, complex I-V responses may appear due to interplay of BTBT and TAT transport mechanisms in these devices.

\section{Conclusions}
\label{sec:conclusions}

In this work, we investigated the transport mechanisms that can occur in APAM pn junctions at cryogenic and room temperature using TCAD simulations. We also explored the possible mechanisms behind the anomalous behavior in the forward-bias response observed in recent measurements\cite{Skeren:2020} at cryogenic temperatures, which deviates from the typical forward-bias response of an Esaki-type diode. These anomalous behaviors include a very low valley voltage and a current suppression at low voltages in the forward-bias response at cryogenic temperatures. We found that the low valley voltage can be the result of a significantly reduced band gap due to the strong confinement of dopants in the $\delta$-layers, which induces a modification of the electronic band structure. We propose a band-gap narrowing (BGN) model that can account for this effect within conventional TCAD simulations. We also found that using a combination of BTBT parameters, we can approximate the current suppression and the NDC behavior. This indicates that there may be two distinct BTBT regimes at work, which can stem from band quantization and realignment due to strong potential confinement in the $\delta$-layers. Finally, we predicted that current leakage and TAT mediated by phonon interactions might play an important role at room temperature. As APAM pn devices become more mature to operate at room temperature, complex I-V responses may appear due to interplay of BTBT, TAT and current leakage in these devices.

\begin{acknowledgments}
This work was supported by the LPS Quantum Collaboratory and by the Laboratory Directed Research and Development (LDRD) program at Sandia National Laboratories. In addition, the work at 3D Epitaxial Technologies was supported in part by U.S. Department of Energy under award DE-SC0020772. Sandia National Laboratories is a multimission laboratory managed and operated by National Technology and Engineering Solutions of Sandia, LLC., a wholly owned subsidiary of Honeywell International, Inc., for the U.S. Department of Energy’s National Nuclear Security Administration under contract DE-NA-0003525. This paper describes objective technical results and analysis. Any subjective views or opinions that might be expressed in the paper do not necessarily represent the views of the U.S. Department of Energy or the United States Government. 
\end{acknowledgments}

\section*{AUTHOR DECLARATIONS}

\subsection*{Conflict of Interest} 
The authors have no conflicts to disclose.

\section*{DATA AVAILABILITY}
The data that support the findings of this study are available from the corresponding author upon reasonable request

\bibliography{main}

\appendix

\section{Trap-assisted tunneling (TAT) model}\label{appendix: TAT model}
For traps not located in the middle of the band gap, the recombination rate \cite{Gao:2019} can be expressed as
\begin{equation}
R_{tat} = \sum_j \frac{np-n_{ie}^2}{\tau_p^j(n+n_t^j)+\tau_n^j(p+p_t^j)} 
\label{eq:tat-eq-1}
\end{equation}
where the summation runs over the total number of different types of traps, $n_{ie}$ is the effective intrinsic concentration of the material, $n_t^j$ and $p_t^j$ for the j-th type of trap are equal to
\begin{equation}
n_t^j=N_C \exp{\bigg( -\frac{E_t^j}{k_BT} \bigg)},~p_t^j=N_V\exp{\bigg( -\frac{E_g-E_t^j}{k_BT} \bigg)}
\label{eq:tat-eq-2}
\end{equation}
where $N_C$ and $N_V$ are the effective density of states in the conduction and valence bands, respectively, $E_t^j$ is the j-th trap energy measured from the conduction band edge, $E_g$ is the effective band gap, $k_B$ is the Boltzmann constant, and $T$ is the lattice temperature. The lifetimes $\tau_n^j$ and $\tau_p^j$ depend on the lattice temperature and the electric field
\begin{equation}
\tau_n^j(T,F)=\frac{\tau_{n0}^j}{1+g_n^j(T,F)},~\tau_p^j(T,F)=\frac{\tau_{p0}^j}{1+g_p^j(T,F)}
\label{eq:tat-eq-3}
\end{equation}
where $\tau_{n0}^j$ and $\tau_{p0}^j$ are field-independent lifetimes, and $g_n^j(T,F)$ and $g_p^j(T,F)$ are the electron and hole field enhancement factors, which capture the recombination enhancement due to band-to-trap tunneling.

\end{document}